\documentclass[prb,aps,showpacs,twocolumn]{revtex4}
\usepackage{dcolumn}
\usepackage{bm}
\usepackage{amsmath}
\usepackage{graphics}

\newcommand{\ket}[1]{|#1\rangle}
\newcommand{\bra}[1]{\langle #1|}
\newcommand{\myint}[1]{\int d#1\,}

\begin{document}

\title{Localized helium excitations in $^4$He$_N$-benzene clusters}

\author{Patrick Huang}
\author{K. Birgitta Whaley}
\email{whaley@socrates.berkeley.edu}
\affiliation{Department of Chemistry and Kenneth S. Pitzer Center for
Theoretical Chemistry, University of California, Berkeley, CA 94720,
USA}

\date{\today}

\begin{abstract}

We compute ground and excited state properties of small helium
clusters $^4$He$_N$ containing a single benzene impurity molecule.
Ground-state structures and energies are obtained for $N=1,2,3,14$
from importance-sampled, rigid-body diffusion Monte Carlo (DMC).
Excited state energies due to helium vibrational motion near the
molecule surface are evaluated using the projection operator,
imaginary time spectral evolution (POITSE) method.  We find excitation
energies of up to $\sim 23$~K above the ground state.  These states
all possess vibrational character of helium atoms in a highly
anisotropic potential due to the aromatic molecule, and can be
categorized in terms of localized and collective vibrational modes.
These results appear to provide precursors for a transition from
localized to collective helium excitations at molecular nanosubstrates
of increasing size.  We discuss the implications of these results for
analysis of anomalous spectral features in recent spectroscopic
studies of large aromatic molecules in helium clusters.

\end{abstract}

\pacs{67.40.Db, 67.70.+n, 36.40.Mr}

\maketitle

\section{Introduction}

Helium droplets provide a unique, ultra-cold nanolaboratory for
investigation of a variety of physical and chemical
phenomena.\cite{toennies01} This has been increasingly used in recent
years to analyze the behavior of a wide variety of atomic and
molecular species in a quantum liquid environment, using spectroscopic
techniques to probe both the molecular and helium
dynamics.\cite{specialJCP} Electronic spectroscopy in particular
allows one to probe microscopic details of the helium dynamics.  In
the large droplet regime ($N\ge 10^3$), the laser-induced fluorescence
(LIF) spectra of molecules in cold helium droplets are usually
characterized by a sharp zero-phonon line (ZPL) due to the transition
for the electronic origin, accompanied by a broad, phonon wing
sideband.  The zero-phonon line can contain fine structure due to
internal molecular transitions, while the phonon wing structure
reflects the coupling to collective excitations of the surrounding
helium.  Thus, electronic excitation spectrum of glyoxal, a small
6-atom molecule (C$_2$H$_2$O$_2$), in $^4$He droplets at $T=0.38$~K,
exhibits a distinct phonon wing feature that is separated by a gap
from the zero-phonon transition.\cite{hartmann96} In contrast, in
$^3$He droplets the corresponding spectrum shows a phonon wing feature
but no gap.\cite{grebenev00a} The phonon wing structure for glyoxal in
has been successfully interpreted in terms of excitation of the
collective phonon-roton modes in large $^4$He
droplets,\cite{hartmann96} while the lack of a gap between zero phonon
and phonon wing features in $^3$He droplets has been interpreted as
consistent with the presence of particle-hole excitations in normal
$^3$He.\cite{grebenev00a} Recent spectroscopic experiments involving
larger organic molecules in $^4$He$_N$ clusters have revealed
interesting additional features beyond these basic
elements.\cite{stienkemeier01} For a number of the larger organic
molecules studied so far, both the ZPL and phonon wing portions of the
LIF spectra exhibit sharp peaks superimposed on the underlying
features.  These additional peaks are not compatible with this basic
picture of molecular electronic excitation coupling to either internal
molecular transitions or bulk compressional modes of
helium.\cite{hartmann98,hartmann01,lindinger01,hartmann02}

In addition to these experiments in large helium droplets, a new class
of size-selective experiments involving small numbers ($N\le 20$) of
$^4$He atoms attached to large planar aromatic molecules has also
recently emerged.\cite{even00,even01} These small cluster studies
allow one to directly observe the size evolution of excited states
involving helium motion, at sizes less than a full solvation shell
around the molecule.  For these small clusters, which are better
described as weakly bound complexes with helium than as quantum
solvated molecules, a very different situation pertains.  Starting
with $N=1$, the experimental spectra show discrete lines that
generally increase in complexity with increasing $N$, with the number
of observed lines reaching a maximum at $N\sim 4-5$.  After this, many
of the discrete features observed for smaller $N$ disappear, until
only a few lines persist at $N\sim 10$.

The larger organic molecules that have been studied experimentally
vary in complexity from planar aromatics such as tetracene (a fused
conjugated system of four six-membered carbon rings connected by
common bonds)\cite{hartmann98,hartmann01} to more complex
heterostructures such as phthalocyanine\cite{hartmann02} and indole
derivatives.\cite{lindinger01} The presence of aromatic character due
to $\pi$-electron conjugation provides a common feature in these
systems.  Because of their geometry and $\pi$-electron character,
planar aromatics such as tetracene are particularly interesting for
study in helium droplets because they can be considered as nanoscale
precursors to a bulk graphite surface and their local quantum
solvation structure concomitantly as a nanoscale precursor of the
adsorption behavior of thin helium films on graphite.  A considerable
body of literature has been accumulated for helium films on
graphite,\cite{bretz73,shirron91,greywall91,zimmerli92,adams92,clements93,wagner94,dalfovo95,csathy98,niyeki98,pierce99}
so that analysis of the solvation structure and helium excitations as
a function of increasing size of polyaromatic molecule offers the
possibility of developing a microscopic understanding of the evolution
from quantum solvation of an isolated molecule, to adsorption and film
formation at a bulk surface in superfluid helium.

The unusual features recorded in experimental spectra for these
organic molecules consist of unidentified peaks in the phonon wings of
vibronic bands\cite{hartmann02} and anomalous splittings of the
zero-phonon lines.\cite{hartmann98,hartmann01,lindinger01} It has been
speculated that both of these types of features may be due to some
type of excitation of helium atoms that are localized at the molecular
surface,\cite{hartmann01,lindinger01,hartmann02} but the true origin
of both of these types of features is unclear.  Supporting evidence
for excitations of localized helium atoms derives from the similarity
of the energies for phonon wing peaks with low-frequency modes of thin
films on graphite ($4-11$~K) observed via neutron scattering
\cite{lauter92,clements96a} as well as from theoretical predictions of
spatially localized helium atoms at an aromatic ring in the benzene
molecule.\cite{kwon01} Path integral calculations reveal this to be a
true localization of helium atoms that are effectively completely
removed from participation in the superfluid, thereby constituting a
single "dead" atom in the surrounding solvation layer.  The
characteristics of excitations of such localized helium atoms around
aromatic substrates are expected to be very different from the
collective helium modes found in large
droplets,\cite{ramakrishna90,chin92,chin95b} but to show increasing
similarity with surface localized modes in helium
films\cite{lauter92,clements96a} as the size of the molecular
nanosubstrate increases.

Theoretical analysis of the phenomenology of these helium
excitations near a molecular nanosubstrate is rendered difficult due
both to the lack of accurate interaction potentials for these larger
molecules with helium, and to the need for accurate calculation of
excited states in a very inhomogeneous quantum liquid environment.
Only a few direct calculations to elucidate the nature of these helium
excitations in the presence of a large molecular impurity have been
reported so far, and these have been restricted to very small numbers
of helium atoms ($N = 1, 2$).\cite{bach97,anderson00,heidenreich01}
Calculations involving doped clusters require accurate helium-impurity
potential energy surfaces, of which there is a definite lack for the
larger organic molecules.  Thus, most calculations with large
molecules resort to the use of simple atom-atom pair potential models.
Given a potential surface, excited state calculations employing basis
set methods are still limited to small $N$, where $N$ is typically no
greater than two.\cite{bach97,anderson00,heidenreich01} For larger
numbers of helium atoms that are however not yet in the regime of bulk
films, a practical approach is provided by calculations for excited
states that are based on quantum Monte Carlo methods.

In this work, we present a series of such quantum Monte Carlo studies
of the ground and excited states of small helium clusters
$(N=1,2,3,14)$ containing a benzene impurity.  Benzene is a simple
planar aromatic molecule, and it can be viewed as a primitive
$sp^3$-hybridized unit of a bulk graphite surface.  Thus, it is the
logical starting point for the kind of systematic study of helium
adsorption on molecular nanosubstrates mentioned above.  We note that
this notion has been pursued extensively for the heavier noble
gases,\cite{leutwyler87} but it is only recently that experimental
data for helium has become available.  We employ here both the
well-known ground-state diffusion Monte Carlo method and the
projection operator imaginary time spectral evolution (POITSE) method
for excited states.\cite{blume97} The latter approach allows exact
excitation energies to be obtained, provided a high-quality trial
function approximating the ground state is available.  We calculate a
range of helium excitation energies at $T=0$~K and analyze these in
terms of their spatial nature and extent.  The previous path integral
calculations for the $^4$He$_{39}$-benzene system at a temperature of
$T=0.625$~K have shown that the effect of the strength and strong
$\pi$-anisotropy of the He-benzene interaction serves to localize a
single helium atom on each side of the benzene surface.\cite{kwon01}
That is, in the Feynman path integral representation, a single helium
atom attached to the benzene surface is completely removed from
permutation exchanges with the surrounding helium environment.  In
this work, we find a set of collective helium vibrations which have
energies of up to $\sim 23$~K above the ground state, and can be
characterized in terms of their transformation properties under the
symmetry group of the He-benzene interaction potential.  We show that
even for sizes as small as $N=14$ helium atoms, one can distinguish a
subset of higher-energy excitations localized near the global minimum
of the helium-benzene interaction potential, from a subset of
lower-energy collective excitations which are delocalized around the
periphery of the molecule surface.  The former appear to be associated
with the localized helium density identified from path integral
calculations in Ref.~\onlinecite{kwon01}.

Sec.~\ref{sec:model} begins with a discussion of the model Hamiltonian
and potential surface.  Technical details of the quantum Monte Carlo
methodology are presented in Sec.~\ref{sec:methods}.  Our results for
ground and excited states of $^4$He$_N$-benzene ($N=1,2,3,14$) are
presented in Sec.~\ref{sec:results}, where we analyze the nature of
these molecule-induced localized states around the benzene impurity
and, and in Sec.~\ref{sec:conclusions} we discuss the implications for
helium excitations on larger aromatic molecules.

\section{\label{sec:model}The model Hamiltonian}

We treat the $^4$He$_N$-benzene cluster as a collection of $N$
indistinguishable helium atoms, and a rigid benzene molecule which is
free to translate and rotate in space.  Thus the benzene
intramolecular degrees of freedom are held fixed, which implicitly
assumes a separation between the helium motion and the benzene
vibrational modes.  The positions of the $N$ helium atoms in a
space-fixed (SF) frame of reference are denoted as
$\{\mathbf{R}_1,\mathbf{R}_2,\ldots,\mathbf{R}_N\}$, and the SF
position of the benzene center-of-mass is denoted as $\mathbf{R}_I$.
The Hamiltonian $\hat{H}$ is $(3N+6)$--dimensional: there are $3N$
helium translational degrees of freedom, plus an additional six
dimensions due to translation and rigid-body rotation of the benzene
molecule.  It takes the form
\begin{equation}
\hat{H} = \hat{T}^{(\mathrm{imp})} + \hat{T}^{(\mathrm{He})} + \hat{V}, 
\label{eq:hamil}
\end{equation}
where $\hat{T}^{(\mathrm{imp})}$, $\hat{T}^{(\mathrm{He})}$, and
$\hat{V}$ are the impurity rigid-body kinetic energy, helium kinetic
energy, and total potential energy, respectively.  The impurity
kinetic energy is most easily expressed by introducing a moving
body-fixed (BF) frame $(\hat{x},\hat{y},\hat{z})$, whose origin
relative to the SF frame is fixed at the benzene center-of-mass ${\bf
R}_I$.  The Euler angles $(\varphi,\vartheta,\chi)$ specify the
orientation of the BF axes, which are set parallel to the benzene
principal axes.  The benzene kinetic energy is thus
\begin{equation}
\hat{T}^{(\mathrm{imp})} = -D_I\nabla_I^2 - 
d_a\left(\frac{\partial^2}{\partial\phi_x^2} + 
\frac{\partial^2}{\partial\phi_y^2}\right) - 
d_c\frac{\partial^2}{\partial\phi_z^2}, \label{eq:impke}
\end{equation}
with the prefactors
\begin{equation}
D_I=\frac{\hbar^2}{2m_I},\ d_a=\frac{\hbar^2}{2I_a},\ d_c=\frac{\hbar^2}{2I_c}. 
\label{eq:constants}
\end{equation}
The first term of $\hat{T}^{(\mathrm{imp})}$ is the center-of-mass
translational kinetic energy for a benzene molecule, where we use a
mass $m_I=78.114$~amu.  The remaining terms give the benzene
rigid-body rotational kinetic energy, where the prefactors
$d_a=0.188$~cm$^{-1}$ and $d_c=0.0938$~cm$^{-1}$ are the (oblate)
symmetric top rotational constants.  The Laplacian $\nabla_I^2$ is
taken with respect to the translations of the benzene center-of-mass.
The angular second derivatives are taken with respect to rotations of
the benzene about the BF axes $(\hat{x},\hat{y},\hat{z})$.  Similarly,
the next term of the Hamiltonian is the helium kinetic energy,
\begin{equation}
\hat{T}^{(\mathrm{He})} = -D_4\sum_{j=1}^N\nabla_j^2,\ D_4=\frac{\hbar^2}{2m_4},
\end{equation}
where $m_4$ is the $^4$He mass, and $\nabla_j^2$ denotes the Laplacian
taken with respect to the displacement of helium atom $j$.

\begin{figure}
\resizebox{\columnwidth}{!}{\includegraphics{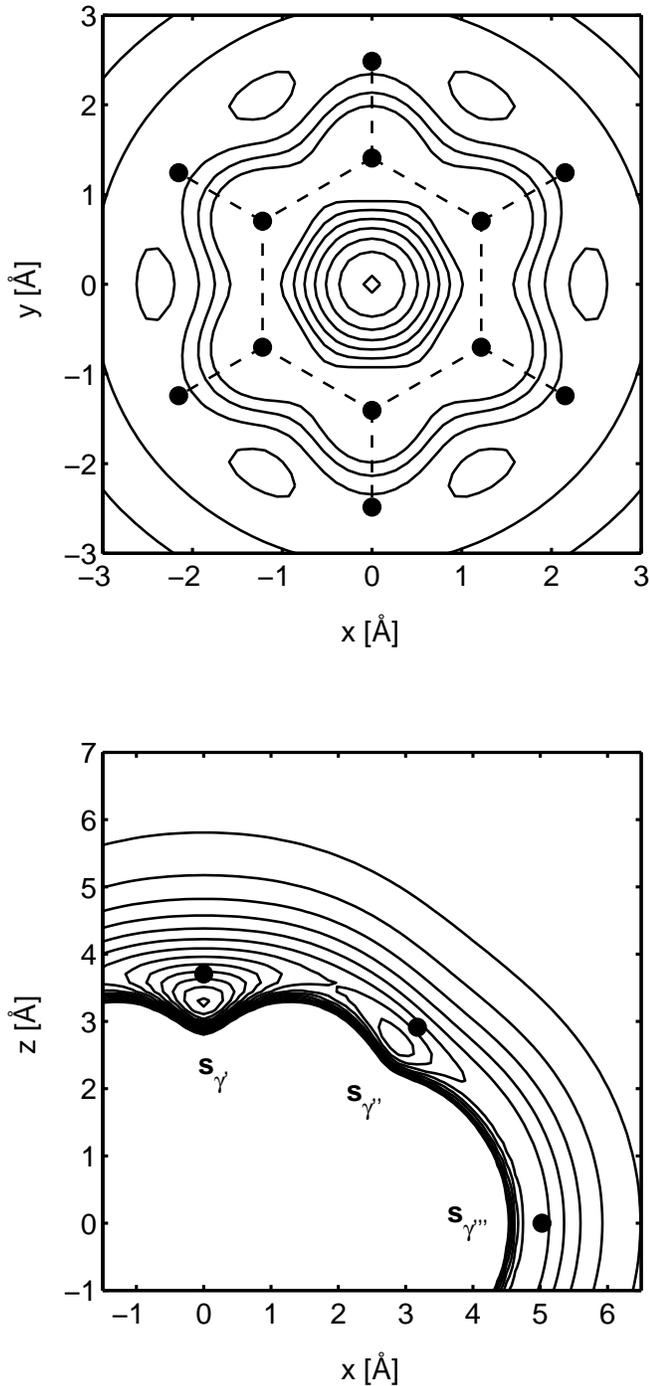}}
\caption{\label{fig:heben}Top: cut of the $^4$He-benzene potential
through the global minima at $z=3.27$~\AA\@.  The carbon and hydrogen
$(x,y)$ positions are marked with solid circles and connected by
dashed lines.  Bottom: cut of the same potential at $y=0$.  Three
representative sites
$\mathbf{s}_{\gamma'},\mathbf{s}_{\gamma''},\mathbf{s}_{\gamma'''}$
are marked, which correspond to centers of the helium-benzene
localization factors centered near the global, secondary, and tertiary
potential minima, respectively (Secs.~\ref{sec:trialfunc} and
\ref{subsec:groundstate}).}
\end{figure}

The final term of Eq.~(\ref{eq:hamil}) is the model potential energy.
We use an additive, two-body potential,
\begin{equation}
\hat{V} = \sum_{j=1}^N V_{\mathrm{He-I}}(\mathbf{r}_j) + \sum_{i<j}^N 
V_{\mathrm{He-He}}(r_{ij}).
\end{equation}
The helium-helium interaction $V_{\mathrm{He-He}}$ is given by the
semi-empirical HFD-B potential of Aziz {\em et al.},\cite{aziz87} and
depends only on the distance $r_{ij}$ between helium atoms $i,j$.  The
helium-benzene interaction $V_{\mathrm{He-I}}$ is most conveniently
expressed in terms of helium BF frame coordinates $\{\mathbf{r}_1,{\bf
r}_2,\ldots,\mathbf{r}_N\}$.  It is an analytical fit to {\em ab
initio} MP2 data of Hobza {\em et al.},\cite{hobza92} and consists of
a sum of atom-atom terms:
\begin{equation}
V_{\mathrm{He-I}}(\mathbf{r}_j) = \sum_{\alpha=1}^6 
V_{\mathrm{C}}(\mathbf{r}_{\alpha j}) + \sum_{\beta=1}^6 V_{\mathrm{H}}(r_{\beta 
j}). \label{eq:hebenzene}
\end{equation}
The indices $\alpha$ and $\beta$ run over the carbon and hydrogen
atoms of the benzene molecule, respectively.  The helium-hydrogen
interaction $V_{\mathrm{H}}$ is a standard Lennard-Jones 6-12 form
which depends on the distance $r_{\beta j}$ between hydrogen atom
$\beta$ and helium atom $j$,
\begin{equation}
V_{\mathrm{H}}(r_{\beta j}) = 4\epsilon_{\mathrm{H}} 
\left[\left(\frac{\sigma_{\mathrm{H}}}{r_{\beta j}}\right)^{12} - 
\left(\frac{\sigma_{\mathrm{H}}}{r_{\beta j}}\right)^6 \right].
\end{equation}
The helium-carbon interaction $V_{\mathrm{C}}$ has a modified
angle-dependent Lennard-Jones 8-14 form,
\begin{equation}
V_{\mathrm{C}}(\mathbf{r}_{\alpha j}) = 4\epsilon'_{\mathrm{C}} 
\left[\left(\frac{\sigma'_{\mathrm{C}}}{r_{\alpha j}}\right)^{14} - 
\left(\frac{\sigma'_{\mathrm{C}}}{r_{\alpha j}}\right)^8 \right],
\end{equation}
where $\epsilon'_{\mathrm{C}} = \epsilon_{\mathrm{C}}\cos^2\theta$,
$\sigma'_{\mathrm{C}} = \sigma_{\mathrm{C}}\cos\theta$, and $\theta$
is the spherical polar angle with respect to a coordinate frame
centered on the carbon atom $\alpha$ and parallel to the BF frame.
The potential fit parameters are listed in Table~\ref{tab:poten}.
\begin{table}
\caption{\label{tab:poten}Helium-benzene potential parameters.  The
carbon-carbon and carbon-hydrogen bond lengths are held fixed at
$r_{\mathrm{CC}}$ and $r_{\mathrm{CH}}$, respectively, and all bond
angles are 120$^{\circ}$.}
\begin{ruledtabular}
\begin{tabular}{cd}
$\epsilon_{\mathrm{C}}$ &       15.25~$K$ \\
$\sigma_{\mathrm{C}}$ &         3.59~$\AA$ \\
$\epsilon_{\mathrm{H}}$ &       19.14~$K$ \\
$\sigma_{\mathrm{H}}$ &		2.69~$\AA$ \\
$r_{\mathrm{CC}}$ &             1.406~$\AA$ \\
$r_{\mathrm{CH}}$ &             1.08~$\AA$
\end{tabular}
\end{ruledtabular}
\end{table}
This helium-benzene potential possesses two equivalent global minima
of $-94.97$~K at $z=\pm 3.27$~\AA\ along the benzene $C_6$-axis.
There are also 12 equivalent secondary minima of $-62.54$~K located
between neighboring hydrogen atoms, six above and six below the
benzene plane.  The global minima are connected to the secondary
minima by saddle points of $-50.77$~K, situated approximately above
and below the C--H bonds.  A cut of the potential at $z=3.27$~\AA\ is
shown in the top panel of Fig.~\ref{fig:heben},
where the $(x,y)$ positions of the carbon and hydrogen atoms on the
$z=0$ plane are also marked.  The lower panel of Fig.~\ref{fig:heben}
shows a cut of the potential along the $y=0$ plane, which reveals one
of the secondary minima at $x=2.84$~\AA, $z=2.70$~\AA, and one of the
saddle points at $x=1.97$~\AA, $z=3.54$~\AA.

For $N=1$, the model Hamiltonian of Eq.~(\ref{eq:hamil}) belongs to
the molecular symmetry group $D_{6h}(M)$, which is the subgroup
consisting of only the feasible elements of the complete nuclear
permutation and inversion group (CNPI).\cite{bunker98} This group is
isomorphic to the $D_{6h}$ point group, which is valid in the limit of
small amplitude helium motions.  For $N>1$, the symmetry group is
identically $D_{6h}(M)$ due to Bose symmetry, since the only allowed
rovibrational states are those obtained from the tensor product of
irreducible representations of $D_{6h}(M)$, with irreducible
representations of the symmetric group $S_N$ which remain unchanged
under permutation of identical $^4$He nuclei, {\em i.e.}\ the totally
symmetric representation of $S_N$.

\section{\label{sec:methods}Computational methodology}

In this section we present a summary of the numerical methods employed
in our study of the $^4$He$_N$-benzene system.  Variational estimates
of ground-state energies are obtained from variational Monte Carlo
(VMC), and exact ground-state properties are computed from diffusion
Monte Carlo (DMC) with importance sampling.\cite{hammond94,viel01} The
projection operator, imaginary time spectral evolution (POITSE)
approach is used to obtain excited-state energies.\cite{blume97} Both
the DMC and POITSE methods can provide results that are exact, to
within a systematic time step error.

\subsection{\label{subsec:vmc}Ground states: variational Monte Carlo (VMC)}

The starting point for our study of the $^4$He$_N$-benzene system
begins with the Rayleigh-Ritz variational theorem,
\begin{equation}
E_T = \frac{\bra{\Psi_T}\hat{H}\ket{\Psi_T}}{\langle\Psi_T|\Psi_T\rangle} \geq 
E_0,
\end{equation}
where the trial energy $E_T$ with respect to some parameterized trial
function $\Psi_T$ represents an upper bound to the exact ground-state
energy $E_0$.  In the coordinate representation, this becomes
\begin{equation}
E_T = \frac{\myint{\mathcal{R}} \Psi_T^2(\mathcal{R}) 
E_L(\mathcal{R})}{\myint{\mathcal{R}} \Psi_T^2(\mathcal{R})}, 
\label{eq:variational}
\end{equation}
where $\mathcal{R}$ is a $(3N+6)$--dimensional coordinate denoting
the positions and orientations of all bodies governed by the
Hamiltonian $\hat{H}$.  The quantity $E_L$ is a local energy, defined
as
\begin{equation}
E_L(\mathcal{R}) = \Psi_T^{-1}(\mathcal{R})\hat{H}\Psi_T(\mathcal{R}).
\end{equation}
An optimized variational upper bound is obtained by varying the
parameters of $\Psi_T$ to minimize $E_T$.

For a realistic $N$-particle helium system, Monte Carlo methods offer
a practical means to compute the multidimensional integral of
Eq.~(\ref{eq:variational}).  In variational Monte Carlo, expectation
values of observables are evaluated as averages over the normalized
distribution
\begin{equation}
f(\mathcal{R}) = \frac{\Psi_T^2(\mathcal{R})}{\myint{\mathcal{R}} 
\Psi_T^2(\mathcal{R})}, \label{eq:fvmc}
\end{equation}
which is numerically represented as a discrete ensemble of Monte Carlo
walkers $\{\mathcal{R}_k\}$.  Typically, we use an ensemble size of
$n_w=2000$ walkers.  This distribution can be generated from a
Metropolis walk, in which a move $\mathcal{R}\rightarrow\mathcal{R}'$
is proposed with a transition probability consisting of factors having
the form
\begin{multline}
G_D(\mathbf{R}_j\rightarrow\mathbf{R}_j',\Delta\tau) = \\
\left(\frac{1}{4\pi D\Delta\tau}\right)^{3/2} \exp\left\{-\frac{[\mathbf{R}_j'-
\mathbf{R}_j-D\Delta\tau\mathbf{F}_j(\mathcal{R})]^2}{4D\Delta\tau}\right\}. 
\label{eq:gdiff}
\end{multline}
The vector quantity $\mathbf{F}$ is commonly referred to as the
``quantum force'', and is chosen to be
\begin{equation}
\mathbf{F}_j(\mathcal{R}) = 2\Psi_T^{-1}(\mathcal{R}) \nabla_j 
\Psi_T(\mathcal{R}).
\end{equation}
$G_D$ can be viewed as the Green's function associated with a
diffusion/drift process in the presence of an external field
$\mathbf{F}$, over a time step of $\Delta\tau$.  All VMC-based
computations reported here use a time step of
$\Delta\tau=75-100$~Hartree$^{-1}$.  For the $^4$He$_N$-benzene system
the full $(3N+6)$--dimensional transition probability is comprised of
$N$ such factors with diffusion constant $D=D_4$ corresponding to the
$N$ helium atoms, a factor with $D=D_I$ corresponding to the benzene
impurity, and three analogous one-dimensional factors corresponding to
individual rotations of the benzene about each of its three principal
axes.  A proposed move $\mathcal{R}\rightarrow\mathcal{R}'$ is
subsequently accepted with probability $p=\min(1,q)$, where the
acceptance ratio $q$ is
\begin{equation}
q = \frac{\Psi_T^2(\mathcal{R}')}{\Psi_T^2(\mathcal{R})} 
\frac{G_D(\mathcal{R}'\rightarrow\mathcal{R},\Delta\tau)}{G_D(\mathcal{R}\rightarrow\mathcal{R}',\Delta\tau)}. \label{eq:metro}
\end{equation}

With the choice of transition and acceptance probabilities given by
Eqs.~(\ref{eq:gdiff})--(\ref{eq:metro}), the Metropolis walk converges
to the asymptotic distribution $\Psi_T^2(\mathcal{R})$.  At this
point, expectation values for quantities such as the variational trial
energy $E_T$ are sampled from this distribution as
\begin{equation}
E_T = \langle E_L \rangle = \left\langle \frac{1}{n_w} \sum_{k=1}^{n_w} 
E_L(\mathcal{R}_k) \right\rangle_{\mathrm{walk}},
\end{equation}
where the notation $\langle\ldots\rangle_{\mathrm{walk}}$ denotes a
statistical average over the course of the equilibrated Metropolis
walk.  This is performed by sampling the quantity given in
$\langle\ldots\rangle_{\mathrm{walk}}$ every $M$ time steps apart,
with $M$ chosen to be longer than the autocorrelation length in order
to minimize correlation biases in the error estimates; typically
$M=50$ for our VMC computations here.

\subsection{\label{subsec:dmc}Ground states: diffusion Monte Carlo (DMC)}

With diffusion Monte Carlo (DMC), one can move beyond the variational
level of theory to obtain {\em exact} ground-state expectation values
for the energy and quantities which commute with the Hamiltonian
$\hat{H}$.  This method derives from the imaginary time
$(\tau=it/\hbar)$ Schr\"{o}dinger equation,
\begin{equation}
-\frac{\partial\Psi(\mathcal{R},\tau)}{\partial\tau} = (\hat{H}-E_{\mathit 
ref})\Psi(\mathcal{R},\tau), \label{eq:schrod}
\end{equation}
where $\Psi(\mathcal{R},\tau)$ is the many-body wave function, and
$E_{\mathit ref}$ is an arbitrary constant shift in the energy
spectrum.  Importance sampling is introduced by multiplying both sides
of Eq.~(\ref{eq:schrod}) by a trial function $\Psi_T(\mathcal{R})$,
and rewriting in terms of $f(\mathcal{R},\tau)=\Psi_T({\mathcal
R})\Psi(\mathcal{R},\tau)$ to obtain a set of equations having the
form
\begin{multline}
-\frac{\partial f(\mathcal{R},\tau)}{\partial\tau} = -D\nabla_j^2 
f(\mathcal{R},\tau) + D\nabla_j\cdot\mathbf{F}_j(\mathcal{R}) 
f(\mathcal{R},\tau) \\
+ [E_L(\mathcal{R}) - E_{\mathit ref}] f(\mathcal{R},\tau). \label{eq:isschrod}
\end{multline}
The stationary solution is now the normalized ``mixed'' distribution
\begin{equation}
f(\mathcal{R}) = 
\frac{\Psi_T(\mathcal{R})\phi_0(\mathcal{R})}{\myint{\mathcal{R}} 
\Psi_T(\mathcal{R})\phi_0(\mathcal{R})}, \label{eq:fmix}
\end{equation}
where $\phi_0$ is the exact many-body ground-state wave function.  In
DMC, the approximate short-time Green's function $G$ which generates
this distribution is
\begin{equation}
G(\mathcal{R}\rightarrow\mathcal{R}',\Delta\tau) \approx 
G_D(\mathcal{R}\rightarrow\mathcal{R}',\Delta\tau) 
G_B(\mathcal{R}\rightarrow\mathcal{R}',\Delta\tau). \label{eq:gdmc}
\end{equation}
Here, $G_D$ is the diffusion/drift Green's function of
Eq.~(\ref{eq:gdiff}), and $G_B$ has the form
\begin{multline}
G_B(\mathcal{R}\rightarrow\mathcal{R}',\Delta\tau) = \\
\exp\left\{-\left[\frac{E_L(\mathcal{R}) + E_L(\mathcal{R}')}{2} - E_{\mathit 
ref}\right]\Delta\tau\right\}.
\end{multline}
All DMC-based computations here use a time step of
$\Delta\tau=25$~Hartree$^{-1}$.  A proposed move
$\mathcal{R}\rightarrow\mathcal{R}'$ is accepted or rejected according
to Eq.~(\ref{eq:metro}), to ensure that the exact ground state is
sampled in the limit of perfect importance sampling, {\em i.e.}\ when
$\Psi_T(\mathcal{R})=\phi_0(\mathcal{R})$.  For a non-exact trial
function however, the approximate Green's function of
Eq.~(\ref{eq:gdmc}) results in a systematic time step error.

Operationally, the importance-sampled DMC procedure is similar to that
described previously for VMC, except now associated with each walker
$\mathcal{R}_k$ is a cumulative weight $w_k$ due to the action of
$G_B$:
\begin{equation}
w_k = \prod_{n=1}^{n_{\tau}} 
G_B(\mathcal{R}_k\rightarrow\mathcal{R}_k',n\Delta\tau).
\end{equation}
The efficiency of the DMC method can be significantly improved by
replicating walkers with large $w_k$, and destroying walkers with
small $w_k$.  At every $10-50$ time steps, the ensemble is checked for
walkers whose weight exceeds the empirically set bounds $w_{\mathit
min}$ and $w_{\mathit max}$.  A walker $\mathcal{R}_k$ with weight
$w_k>w_{\mathit max}$ is replicated into $n_k=\mathrm{int}(w_k+u)$
walkers, where $u$ is a uniform random number on $[0,1)$.  These new
walkers are then assigned the weight $w_k/n_k$.  A walker ${\mathcal
R}_k$ with weight $w_k<w_{\mathit min}$ is destroyed with probability
$1-w_k$; otherwise its weight is set to unity.  This is to ensure that
the branching scheme, on average, does not artificially alter the
ensemble sum of weights $W_{\mathit tot} = \sum w_k$.  The parameters
$w_{\mathit min}$ and $w_{\mathit max}$ are chosen empirically to
maintain a stable DMC walk with respect to $W_{\mathit tot}$ and the
ensemble size; here we use $w_{\mathit min}=0.1-0.3$ and $w_{\mathit
max}=2.0-2.2$.  Additionally, we also vary the reference energy
$E_{\mathit ref}$ during the course of the walk according to
\begin{equation}
E_{\mathit ref}(\tau+\Delta\tau) = E_{\mathit ref}(\tau) + 
\frac{\eta}{\Delta\tau}\ln\left[\frac{W_{\mathit tot}(\tau)}{W_{\mathit 
tot}(\tau+\Delta\tau)}\right].
\end{equation}
Here, the imaginary-time dependence of the $E_{\mathit ref}$ and
$W_{\mathit tot}$ is explicitly written.  We set $E_{\mathit ref}$
initially to be the starting ensemble average for the local energy
$\langle E_L \rangle$.  The empirical update parameter $\eta$ is
chosen to be as small as possible, typically
$\eta/\Delta\tau=0.0004-0.004$, which gives fluctuations in
$W_{\mathit tot}$ of roughly 5\%.

Once the ensemble converges to the mixed distribution
$\Psi_T(\mathcal{R})\phi_0(\mathcal{R})$, mixed expectation values for
a Hermitian observable $\hat{O}$ can be obtained as
\begin{align}
\langle\hat{O}\rangle_{\mathrm{mix}} & = \left\langle \frac{1}{W_{\mathit 
tot}}\sum_{k=1}^{n_w} w_k \Psi_T^{-1}(\mathcal{R}_k)\hat{O}\Psi_T(\mathcal{R}_k) 
\right\rangle_{\mathrm{walk}} \\
& = \frac{\myint{\mathcal{R}} \Psi_T(\mathcal{R})\phi_0(\mathcal{R}) \Psi_T^{-
1}(\mathcal{R})\hat{O}\Psi_T(\mathcal{R})}{\myint{\mathcal{R}} 
\Psi_T(\mathcal{R})\phi_0(\mathcal{R})} \\
& = \frac{\bra{\phi_0}\hat{O}\ket{\Psi_T}}{\bra{\phi_0}\Psi_T\rangle}.
\end{align}
In our ground-state DMC calculations, this average is computed by
sampling $\hat{O}$ every $M=150$ time steps apart, which is larger
than that used for VMC because successive DMC iterations are more
strongly correlated due to the smaller DMC time steps that we use
here.  In the case where $\phi_0$ is an eigenstate of $\hat{O}$, {\em
i.e.}\ when $\hat{O}$ commutes with the Hamiltonian $\hat{H}$, this
procedure yields exact expectation values.  All ground-state energies
reported in this work are derived from the mixed estimator for the
local energy, and are thus exact.  For quantities which do not commute
with $\hat{H}$, such as the local helium density operator
$\hat{\rho}(\mathbf{r})=\sum_j \delta(\mathbf{r}-\mathbf{r}_j)$, the
mixed estimator is biased by the trial function $\Psi_T$.  But as
pointed out by Liu {\em et al.},\cite{liu74} the asymptotic weight
$w_k(\tau\rightarrow\infty)$ of a walker $\mathcal{R}_k$ is
proportional to $\phi_0(\mathcal{R}_k)/\Psi_T(\mathcal{R}_k)$.  Thus,
in principle this trial function bias can be eliminated by computing
instead the reweighted average
\begin{align}
\langle\hat{O}\rangle & = \frac{\sum_{k=1}^{n_w} 
w_k(\tau)w_{k}(\tau+M'\Delta\tau) \Psi_T^{-
1}(\mathcal{R}_k)\hat{O}\Psi_T(\mathcal{R}_k)}{\sum_{k=1}^{n_w} 
w_k(\tau)w_{k}(\tau+M'\Delta\tau)} \label{eq:desc_weight} \\
& = \cfrac{\myint{\mathcal{R}} \Psi_T(\mathcal{R})\phi_0(\mathcal{R}) \Psi_T^{-
1}(\mathcal{R})\hat{O}\Psi_T(\mathcal{R}) 
\cfrac{\phi_0(\mathcal{R})}{\Psi_T(\mathcal{R})}}{\myint{\mathcal{R}} 
\Psi_T(\mathcal{R})\phi_0(\mathcal{R}) 
\cfrac{\phi_0(\mathcal{R})}{\Psi_T(\mathcal{R})}} \\
& = \bra{\phi_0}\hat{O}\ket{\phi_0},
\end{align}
where $M'$ is taken to be as large as possible, typically $M'=1500$.
When branching processes are incorporated in the DMC, one needs to
take care to keep track of walkers that descended from a walker
$\mathcal{R}_k$ at time $\tau$.  All helium densities reported in this
work ($N\leq 14$) are computed by reweighting walkers $\mathcal{R}_k$
by their descendant weights, {\em i.e.}\ using
Eq.~(\ref{eq:desc_weight}).  However, the statistical noise seen in
the densities computed in this manner is much greater than from
densities obtained using the mixed estimator, and thus this approach
becomes problematic for $^4$He$_N$-benzene at still larger sizes.

\subsection{\label{subsec:poitse}Excited states: projection operator imaginary 
time spectral evolution (POITSE)}

The POITSE approach is a DMC-based method which can provide exact
excited state energies, subject to a systematic time step
bias.\cite{blume97} It begins with the DMC evaluation of the
imaginary-time correlation function
\begin{align}
\tilde{\kappa}(\tau) & = \frac{\bra{\Psi_T} \hat{A} e^{-(\hat{H}-E_{\mathit 
ref})\tau} \hat{A}^{\dagger} \ket{\Psi_T}}{\bra{\Psi_T} e^{-(\hat{H}-E_{\mathit 
ref})\tau} \ket{\Psi_T}} \label{eq:corr} \\
                     & \propto \sum_n 
|\bra{\phi_n}\hat{A}^{\dagger}\ket{\Psi_T}|^2 e^{-(E_n-E_0)\tau} + O(x^2) 
\label{eq:poitsedecay}
\end{align}
where
\begin{equation}
x = \frac{\bra{\Psi_T}\phi_m\rangle}{\bra{\Psi_T}\phi_0\rangle}.
\end{equation}
Here, $\{\phi_n\}$ and $\{E_n\}$ represent a complete set of energy
eigenstates and eigenvalues of the Hamiltonian $\hat{H}$,
respectively, and $\hat{A}^{\dagger}$ is an operator chosen to connect
$\Psi_T$ to the excited state(s) of interest $\phi_n$.  $\Psi_T$
should be a good approximation to the exact ground state $\phi_0$.  To
second order in $x$, $\tilde{\kappa}(\tau)$ is a superposition of
exponential decays whose decay rates correspond to the energy
differences $E_n-E_0$.\cite{blume98,huang01} The VMC procedure of
Sec.~\ref{subsec:vmc} is used to generate an initial ensemble of
$n_w=2000$ walkers distributed according to
$f(\mathcal{R},0)\propto\Psi_T^2(\mathcal{R})$.  This distribution is
then propagated in imaginary time using the DMC procedure outlined in
Sec.~\ref{subsec:dmc}, during which the correlation function
$\tilde{\kappa}(\tau)$ is sampled from the DMC walk as\cite{huang01}
\begin{equation}
\tilde{\kappa}(\tau) = \frac{1}{W_{\mathit tot}(\tau)} \sum_{k'}^{n_w(\tau)} 
\hat{A}^{\dagger}(\mathcal{R}^{(0)}_k) \hat{A}(\mathcal{R}^{(\tau)}_{k'}) 
w_{k'}(\tau). \label{eq:ktau}
\end{equation}
The index $k'$ denotes walkers at time $\tau$ which descended from a
parent walker $k$ at time $\tau=0$, and here we also explicitly
emphasize the imaginary-time dependence of the DMC quantities
$W_{\mathit tot}$, $n_w$, and $w_k$.

An inverse Laplace transform of Eq.~(\ref{eq:poitsedecay}) yields the
spectral function
\begin{equation}
\kappa(E) \propto \sum_n |\bra{\phi_n}\hat{A}^{\dagger}\ket{\Psi_T}|^2 \delta(E-
E_n+E_0), \label{eq:spec}
\end{equation}
whose peak positions give the excited state energies $E_n-E_0$.  The
additive contributions $O(x^2)$ are neglected in Eq.~(\ref{eq:spec}),
since they do not affect the positions of the peaks associated with
$E_n-E_0$, and in practice, their spectral weights are negligible for
a reasonable choice of $\Psi_T$.\cite{blume97,blume98} The numerical
inverse Laplace transform is performed using the Maximum Entropy
Method (MEM), in which the Laplace inversion is formulated as a data
analysis problem in terms of Bayesian probability
theory.\cite{jarrell96} We adopt here the vector notation
$\bm{\kappa}\equiv\{\kappa(E_i)\Delta E\}$ to denote the discrete
spectral function specified at intervals of width $\Delta E$, and
$\tilde{\bm{\kappa}}\equiv\{\tilde{\kappa}(\tau_i)\Delta\tau\}$ to
denote the correlation function sampled at intervals of $\Delta\tau$.
The Bayesian approach begins with the posterior distribution function
\begin{equation}
p(\bm{\kappa}|\tilde{\bm{\kappa}},\alpha,\mathbf{m},I) \propto e^{\alpha S - L}.
\end{equation}
The posterior $p$ represents the probability of obtaining the image
$\bm{\kappa}$, given the Monte Carlo data $\tilde{\bm{\kappa}}$, the
parameter $\alpha$, an initial guess for the image $\mathbf{m}$ (also
referred to as the default model), and any other relevant background
information $I$.  We use a constant value for the default model
$\mathbf{m}$.  The quantity $S$ is the Shannon-Jaynes
information-theoretic entropy,
\begin{equation}
S = \sum_i \left[\kappa_i - m_i - 
\kappa_i\log\left(\frac{\kappa_i}{m_i}\right)\right],
\end{equation}
and $L=\chi^2/2$ is chosen to be the usual Gaussian likelihood.  In
matrix notation, $\chi^2$ takes the form
\begin{equation}
\chi^2 = (\hat{\mathcal{L}}\bm{\kappa}-\tilde{\bm{\kappa}})^T \cdot 
\tilde{\mathbf{C}}^{-1} \cdot (\hat{\mathcal{L}}\bm{\kappa}-\tilde{\bm{\kappa}}), 
\label{eq:chisq}
\end{equation}
where $\hat{\mathcal{L}}$ is the Laplace operator, and
$\tilde{\mathbf{C}}$ is the covariance matrix for the data
$\tilde{\bm{\kappa}}$.  In this work, $\tilde{\mathbf{C}}$ is obtained
from averaging $N_d=256$ decays, which is sufficient to give
well-converged spectra for $N\leq 3$.  A search is then made for the
image $\hat{\bm{\kappa}}$ which maximizes the posterior $p$, using a
search algorithm due to Bryan.\cite{bryan90} In the limit $\alpha=0$,
this reduces to a standard least-squares data-fitting procedure
involving the minimization of $\chi^2$, which is numerically unstable
when one seeks to infer a free-form solution for the image from data
with non-negligible Monte Carlo noise.  For $\alpha>0$, the search for
$\hat{\bm{\kappa}}$ requires that the entropy $S$ be simultaneously
maximized while $\chi^2$ is minimized.  Thus, $\alpha$ can be viewed
as a regularization parameter, which stabilizes the least-squares fit
by constraining the minimization of $\chi^2$.

In the ideal situation where the spectrum consists of a single, sharp
peak, we take the excitation energy $E_n-E_0$ to be the first moment
of the spectrum,
\begin{equation}
E_n-E_0 = \langle E \rangle = \frac{\sum_i E_i \hat{\kappa}_i}{\sum_i 
\hat{\kappa}_i}. \label{eq:mem_eavg}
\end{equation}
The corresponding estimate for the variance of the mean $\sigma^2$
then follows from the usual procedure for the propagation of
uncertainties,
\begin{equation}
\sigma^2 = \left(\frac{\langle E \rangle}{\partial\hat{\bm{\kappa}}}\right)^T 
\cdot \mathbf{C} \cdot \left(\frac{\langle E 
\rangle}{\partial\hat{\bm{\kappa}}}\right), \label{eq:mem_evar}
\end{equation}
which requires knowledge of the covariance matrix $\mathbf{C}$ for the
image $\hat{\bm{\kappa}}$.  By approximating $p$ as a sharply-peaked
Gaussian in the neighborhood of $\hat{\bm{\kappa}}$, the covariance of
the image becomes\cite{gubernatis91}
\begin{equation}
\mathbf{C} = -(\alpha\nabla\nabla S - \nabla\nabla L)^{-1}.
\end{equation}
When the spectrum is composed of multiple, well-separated peaks, the
mean value of the peak and the variance in the mean is obtained in the
same manner as given by Eqs.~(\ref{eq:mem_eavg})--(\ref{eq:mem_evar}),
except that the summation is taken only over the spectral feature of
interest.  Note that the approach taken here in evaluating error
estimates differs somewhat from the Gaussian approach advocated in
Ref.~\onlinecite{sivia96}.  In the Gaussian approach the variance in
$E_n-E_0$ is associated with the peak width, and thus does not
necessarily scale as $1/N_d$, where $N_d$ is the number of DMC decays
used to compute statistics for the covariance matrix
$\tilde{\mathbf{C}}$ of Eq.~(\ref{eq:chisq}).  Here, we obtain
$E_n-E_0$ from the {\em mean} peak position, and then evaluate the
variance of the mean according to
Eqs.~(\ref{eq:mem_eavg})--(\ref{eq:mem_evar}), which by construction
scales as $1/N_d$.

\section{\label{sec:trialfunc}Trial functions and excitation operators}

While in principle a fully converged DMC result should be independent
of the trial function $\Psi_T$, a good choice of $\Psi_T$ which
closely approximates the exact ground state $\phi_0$ can significantly
improve computational efficiency.  On the other hand, a poor choice of
$\Psi_T$ can produce misleading results.  In this study, we use
analytical forms motivated by basic physical considerations.  The
trial function is a product of two-body factors,
\begin{equation}
\Psi_T(\mathcal{R}) = \prod_{j=1}^N \xi(\mathbf{r}_j) \prod_{i<j}^N \chi(r_{ij}). 
\label{eq:psitrial}
\end{equation}
Here, $\chi$ describes helium-helium correlations, and has the
McMillan form\cite{mcmillan65}
\begin{equation}
\chi(r_{ij}) = \exp\left(-\frac{c_{\mathrm{He}}}{r_{ij}^5}\right). 
\label{eq:chitrial}
\end{equation}
The function $\xi$ describes helium-benzene correlations, and is a
product of helium-benzene factors defined in the BF frame,
\begin{equation}
\xi(\mathbf{r}_j) = \prod_{\gamma=0}^{n_{\gamma}} \xi_{\gamma}(\mathbf{r}_j). 
\label{eq:xitrial}
\end{equation}
The first of these factors $\xi_0$ controls the behavior of the trial
function at short and long helium-benzene separations:
\begin{multline}
\xi_0(\mathbf{r}_j) = \exp\left[-\sum_{\alpha=1}^6 
\frac{c_{\mathrm{C}}}{r_{\alpha j}^6} -\sum_{\beta=1}^6 
\frac{c_{\mathrm{H}}}{r_{\beta j}^5} \right. \\
\left. - a_0(x_j^2+y_j^2) - c_0 z_j^2\right]. \label{eq:xi0}
\end{multline}
The terms involving the parameters $c_{\mathrm{C}},c_{\mathrm{H}}$ ensure that 
$\Psi_T$ goes
to zero as a helium atom approaches a carbon or hydrogen atom, where
$r_{\alpha j}$ and $r_{\beta j}$ denote helium-carbon and
helium-hydrogen distances, respectively.  The $r^{-6}$ and $r^{-5}$
forms are chosen to cancel the leading singularity in the local
energy\cite{mushinski94} due to the helium-benzene pair potential of
Eq.~(\ref{eq:hebenzene}).

With $\xi=\xi_0$ alone, it is known that the resulting trial function
describes a ground state with ``liquid''-like characteristics.  If the
true ground state has ``solid''-like qualities, {\em i.e.}\ individual
helium atoms are strongly localized around the molecular impurity,
such a state will not solidify out of a liquid-like trial function at
the variational level of theory, within reasonable VMC simulation
times.  One simple way to incorporate solid-like features {\em a
priori} is to introduce a set of sites $\{\mathbf{s}_{\gamma}\}$ fixed
in the BF frame.  The remaining factors $\xi_{\gamma}\ (\gamma>0)$
serve to localize helium atoms near these sites.  In this work we use
the exponential of a Gaussian to effect this
localization:\cite{reatto79}
\begin{equation}
\xi_{\gamma}(\mathbf{r}_j) = \exp\left(c_{\gamma} e^{-a_{\gamma}|\mathbf{r}_j-
\mathbf{s}_{\gamma}|^2}\right). \label{eq:xigamma}
\end{equation}
The specification of the sites $\{\mathbf{s}_{\gamma}\}$ is given
later in Sec.~\ref{subsec:groundstate}.  A more sophisticated
variational approach would involve the extension of this trial
function to shadow functions, in which the $\{\mathbf{s}_{\gamma}\}$
are treated as subsidiary variables, thus obviating the need for an
{\em a priori} specification of $\{{\bf
s}_{\gamma}\}$.\cite{vitiello88} While a shadow trial function is
expected to yield a significantly better description of the ground
state, we do not employ this approach here because it introduces
additional complexity due to the need to sample the subsidiary
variables $\{\mathbf{s}_{\gamma}\}$ in VMC and DMC.

In the POITSE formulation, an initial {\em ansatz} for the excited
state is generated by the action of an excitation operator
$\hat{A}^{\dagger}$ on $\Psi_T$, and exact excited-state energies are
then extracted from the DMC imaginary-time evolution of this initial
state.  Due to the inherent numerical difficulties in the MEM
inversion when the target image $\kappa(E)$ contains multiple,
closely-spaced peaks of comparable spectral weight, we seek operators
$\hat{A}^{\dagger}$ which connect $\Psi_T$ to only one or a few
energetically well-separated states.  Thus, one consideration is to
choose $\hat{A}^{\dagger}$ to transform as an irreducible
representation of the $D_{6h}(M)$ group, such that the spectral weight
$\bra{\phi_n} \hat{A}^{\dagger} \ket{\Psi_T}$ is only non-zero for
states $\phi_n$ which transform identically.  For systems with a
relatively high degree of symmetry, such a choice of
$\hat{A}^{\dagger}$ would significantly reduce the number of
individual decays contributing to $\tilde{\kappa}(\tau)$.

In the BF representation, a convenient primitive for
$\hat{A}^{\dagger}$ are provided by the regular spherical harmonics:
\begin{equation}
R_{lm}(\mathbf{r}) = r^l \left(\frac{4\pi}{2l+1}\right)^{1/2} 
Y_{lm}(\theta,\phi). \label{eq:rlm}
\end{equation}
Similar operators have been used previously to study rotationally
excited states of small $^4$He$_N$ and (H$_2$)$_N$ clusters
($N=7,40$).\cite{mcmahon93,cheng96b} For a pure cluster, this yields
trial excited states which are simultaneous eigenstates of the square
of the total helium angular momentum $\hat{L}^2$, and its SF
$z$-component $\hat{L}_z$.  Addition of a benzene impurity lowers the
symmetry to $D_{6h}(M)$.  In this case the set of regular spherical
harmonics can be symmetrized by taking linear combinations of $R_{lm}$
and $R_{l,-m}$ to obtain real operators which transform as irreducible
representations of $D_{6h}(M)$:
\begin{align}
 R_{lmc} & = \sqrt{\frac{1}{2}}\left[(-1)^m R_{lm} + R_{l,-m}\right], 
\label{eq:rlmc} \\
iR_{lms} & = \sqrt{\frac{1}{2}}\left[(-1)^m R_{lm} - R_{l,-m}\right], 
\label{eq:rlms}
\end{align}
The subscripts $c$ and $s$ indicate proportionality to $\cos(m\phi)$
and $\sin(m\phi)$, respectively.  The various $R_{lm}$ are listed in
Table~\ref{tab:proj}
\begin{table}
\caption{\label{tab:proj}Symmetrized regular spherical harmonics
$R_{lm}$ of Eq.~(\ref{eq:rlm})--(\ref{eq:rlms}), in terms of BF
Cartesian coordinates (to within a constant real factor).  The last
column gives its corresponding symmetry $\Gamma$ according to the
irreducible representations of the $D_{6h}(M)$ group.}
\begin{ruledtabular}
\begin{tabular}{ccc}
$R_{lm}$ &		&					$\Gamma$ \\
\hline
$R_{10}$ &              $z$ &					$A_{2u}$ \\
$R_{11c},\,R_{11s}$ &	$x,\:y$ &				$E_{1u}$ \\
$R_{20}$ &		$x^2+y^2-2z^2$ &			$A_{1g}$ \\
$R_{21c},\,R_{21s}$ &	$xz,\:yz$ &				$E_{1g}$ \\
$R_{22c},\,R_{22s}$ &	$x^2-y^2,\:xy$ &			$E_{2g}$ \\
$R_{30}$ &		$z(x^2+y^2-\frac{2}{3}z^2)$ &		$A_{2u}$ \\
$R_{31c},\,R_{31s}$ &	$x(x^2+y^2-4z^2),\:y(x^2+y^2-4z^2)$ &	$E_{1u}$ \\
$R_{32c},\,R_{32s}$ &	$z(x^2-y^2),\:xyz$ &			$E_{2u}$ \\
$R_{33c}$ &		$x^3-3xy^2$ &				$B_{2u}$ \\
$R_{33s}$ &		$3x^2y-y^3$ &				$B_{1u}$ \\
$R_{40}$ &		$8z^2[z^2-3(x^2+y^2)]+3(x^2+y^2)^2$ &	$A_{1g}$ \\
$R_{41c},\,R_{41s}$ & $xz[4z^2-3(x^2+y^2)],\:yz[4z^2-3(x^2+y^2)]$ & $E_{1g}$ \\
$R_{42c},\,R_{42s}$ & $(x^2-y^2)(6z^2-x^2-y^2),\:xy(6z^2-x^2-y^2)$ & $E_{2g}$ \\
$R_{43c}$ &		$z(x^3-3xy^2)$ &			$B_{1g}$ \\
$R_{43s}$ &		$z(3x^2y-y^3)$ &			$B_{2g}$ \\
$R_{44c},\,R_{44s}$ &	$(x^2-y^2)^2-(2xy)^2,\:xy(x^2-y^2)$ &	$E_{2g}$ \\
\end{tabular}
\end{ruledtabular}
\end{table}
in terms of BF Cartesian coordinates for up to $l=4$, along with their
corresponding symmetry labels under $D_{6h}(M)$.  Note that for up to
$l=4$ under $D_{6h}(M)$, $R_{lmc}$ and $R_{lms}$ are doubly degenerate
except for $m=3$, and thus from this point on we drop the subscripts
$c$ and $s$ for the degenerate functions except when the distinction
is necessary.  For $N>1$, these one-particle operators are
Bose-symmetrized over the $N$ identical bosons to give
single-excitation operators:
\begin{equation}
\hat{A}^{\dagger}_{lm}(\mathcal{R}) = \sum_{j=1}^N R_{lm}(\mathbf{r}_j).
\end{equation}
Application of $\hat{A}_{lm}^{\dagger}$ on $\Psi_T$ then gives the
trial excited state:
\begin{align}
\Psi & = \hat{A}_{lm}^{\dagger} \Psi_T \\
     & = \sum_{j=1}^N \xi(\mathbf{r}_1) \xi(\mathbf{r}_2) \ldots 
R_{lm}(\mathbf{r}_j) \xi(\mathbf{r}_j) \ldots \xi(\mathbf{r}_N) \prod_{i<j}^N 
\chi(r_{ij}). \label{eq:trial_a}
\end{align}
Thus, the action of $\hat{A}^{\dagger}_{lm}$ is to impose a trial
nodal structure on the helium-benzene factor $\xi$, generating a trial
excited state corresponding to a Bose-symmetrized single-excitation
excited state.  As will be discussed further in
Sec.~\ref{subsec:excite}, these operators can be further refined to
take into account the localization of the helium density near the
molecule surface.  In principle this approach can also be readily
extended to multi-excitation operators.

\section{\label{sec:results}Results and discussion}

\subsection{\label{subsec:groundstate}Ground states}

Ground-state energies and structures for the $^4$He$_N$-benzene
cluster ($N=1,2,3,14$) were computed using the methodologies described
in Secs.~\ref{subsec:vmc} and \ref{subsec:dmc}.  Our general strategy
for ground states is to first begin with a preliminary unbiased DMC
calculation with $\Psi_T=1$, which gives an asymptotic DMC
distribution proportional to the exact Bose ground state $\phi_0$.
However, such an unbiased DMC approach for helium clusters has a
number of convergence issues associated with it.  In the absence of a
guiding function, the DMC walk will spend too much time sampling
unimportant regions of configuration space, and thus unbiased DMC can
give an energy which is not converged and is higher than the true
value of $E_0$.\cite{viel01} Nevertheless, it can be useful in
situations where no other information is available, in particular to
act as a guide for constructing a suitable starting trial function
$\Psi_T$.  We first compute a reduced ground-state wave function
$\tilde{\phi}_0(\mathbf{r})$ in the BF frame by binning the unbiased
DMC distribution $\phi_0(\mathcal{R})$ onto a three-dimensional grid.
Trial function parameters for the helium-benzene factor $\xi$,
Eqs.~(\ref{eq:xitrial})--(\ref{eq:xigamma}), are then obtained by
fitting to this $\tilde{\phi}_0(\mathbf{r})$.  In some cases, these
parameters are further varied to lower the variational energy, and the
resulting trial function is then used as input for an
importance-sampled DMC calculation.  The final optimized trial
function parameters for the various sizes are listed in
Table~\ref{tab:trialparam}.
\begin{table}
\caption{\label{tab:trialparam}Parameters for $\Psi_T$ of
Eqs.~(\ref{eq:psitrial})--(\ref{eq:xigamma}), in atomic units.  The
position of a site $\mathbf{s}_{\gamma}$ is listed in terms of its
$(x,y,z)$ Cartesian coordinates in the BF frame.  Only one center from
a set of sites which are equivalent by symmetry is given, and the
remaining can be generated by application of the elements of the
$D_{6h}(M)$ group.}
\begin{ruledtabular}
\begin{tabular}{cdddd}
$N$ &		1	& 2	 & 3		 & 14 \\
\hline
$c_{\mathrm{He}}$ & 4404.1 & 4404.1 & 4404.1	& 3674.6 \\
$c_{\mathrm{C}}$ &		6000.0	& 6000.0 & 6590.9	& 8217.7 \\
$c_{\mathrm{H}}$ &		8000.0	& 8000.0 & 3519.7	& 2546.2 \\
$a_0$ &		0.05	& 0.05	 & 0.0078624	& 0.014378 \\
$c_0$ &		0.06	& 0.06	 & 0.01441	& 0.0073448 \\
$c_{\gamma'}$ &		&	 & 2.79		& 1.9317 \\
$a_{\gamma'}$ &		&	 & 0.090199	& 0.15613 \\
$\mathbf{s}_{\gamma'}$ &	&	 & \multicolumn{1}{c}{(0.0,0.0,7.0)} & 
\multicolumn{1}{c}{(0.0,0.0,7.0)} \\
$c_{\gamma''}$ &		&	 & 1.7103	& 1.5165 \\
$a_{\gamma''}$ &		&	 & 0.090199	& 0.095327 \\
$\mathbf{s}_{\gamma''}$ &	&	 & \multicolumn{1}{c}{(6.0,0.0,5.5)} & 
\multicolumn{1}{c}{(6.0,0.0,5.5)} \\
$c_{\gamma'''}$ &		&	 &		& 2.217 \\
$a_{\gamma'''}$ &		&	 &		& 0.030121 \\
$\mathbf{s}_{\gamma'''}$ &	&	 &		& 
\multicolumn{1}{c}{(9.5,0.0,0.0)} \\
\end{tabular}
\end{ruledtabular}
\end{table}

The VMC and DMC results for the ground-state energy per particle
$E_0/N$ are summarized in Table~\ref{tab:E0}.
\begin{table}
\caption{\label{tab:E0}Ground-state energy per particle $E_0/N$, in
Kelvins.  The second column from left gives the variational energy
from VMC with respect to the trial function $\Psi_T$, parameterized by
values given in Table~\ref{tab:trialparam}.  Unbiased DMC refers to
DMC with $\Psi_T=1$.  IS-DMC is the best estimate for the exact
ground-state energy, using the trial function whose corresponding VMC
energy is given in the second column.  The value in parenthesis
represents the standard deviation in the last significant figure.}
\begin{ruledtabular}
\begin{tabular}{cddd}
$N$ & \multicolumn{1}{c}{VMC} & \multicolumn{1}{c}{unbiased DMC} & 
\multicolumn{1}{c}{IS-DMC} \\
\hline
1 &	-1.9(2) &	-26.72(7) &		-26.83(4) \\
2 &	-4.0(2) &	-27.0(1) &		-27.12(3) \\
3 &	-9.6(2) &	-23.8(1) &		-23.85(2) \\
14 &	-9.6(3) &	-16.11(5) &		-17.51(1)
\end{tabular}
\end{ruledtabular}
\end{table}
For $N=1-2$, we find that a simple trial function involving
$\xi=\xi_0$ only, Eq.~(\ref{eq:xi0}), is sufficient to give good DMC
results.  With the two global minima occupied at $N=2$, additional
helium atoms begin to occupy the twelve secondary potential minima
(Fig.~\ref{fig:heben}).  To describe this situation for $N=3$, we
incorporate localization factors $\xi_{\gamma}$,
Eq.~(\ref{eq:xigamma}), centered at sites $\{\mathbf{s}_{\gamma'}\}$
and $\{\mathbf{s}_{\gamma''}\}$.  The set of sites
$\{\mathbf{s}_{\gamma'}\}$ are centered near the two equivalent global
minima of the helium-benzene potential, the set
$\{\mathbf{s}_{\gamma'}\}$ are centered near the twelve equivalent
secondary potential minima.  The $N=14$ trial function is determined
in the same manner as $N=3$, except that we now also incorporate
additional localization factors centered at sites
$\{\mathbf{s}_{\gamma'''}\}$ which are situated near the six
equivalent tertiary potential minima.  This allows for a description
of helium binding in these regions in the larger clusters.  Three
representatives of the sites
$\mathbf{s}_{\gamma'},\mathbf{s}_{\gamma''},\mathbf{s}_{\gamma'''}$
are marked on the potential contour plot of Fig.~\ref{fig:heben}.  The
positions of all corresponding sites are readily generated from
application of the elements of the $D_{6h}(M)$ group.

\begin{figure*}
\resizebox{\textwidth}{!}{\rotatebox{-90}{\includegraphics{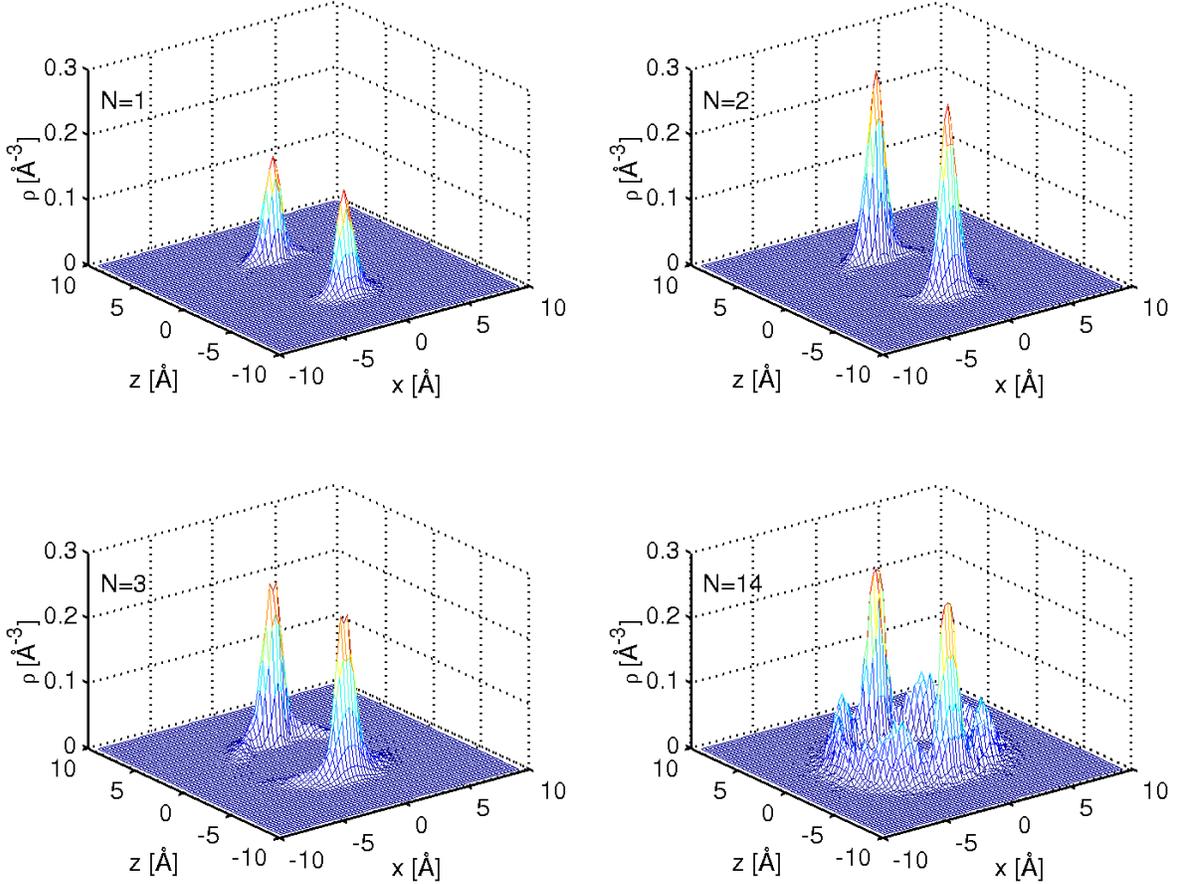}}}
\caption{\label{fig:meshden}Cuts of the helium density in the BF frame
along the $y=0$ plane, for $N=1,2,3,14$.  The two dominant peaks at
$z=\pm 3.7$~\AA\ each correspond to a single helium atom situated on
the molecule $C_6$-axis, above and below the benzene plane.}
\end{figure*}

Cuts of the helium density along the $y=0$ plane are shown in
Fig.~\ref{fig:meshden}.
These cuts are taken along a plane perpendicular to the benzene
molecular plane, bisecting two parallel carbon-carbon bonds.  All
structural quantities given here are obtained from DMC using the
descendant weighting procedure described in Sec.~\ref{subsec:dmc}, and
are thus sampled from the exact ground-state density distribution
$\phi_0^2$.  For $N=1,2$, the helium density along the benzene
$C_6$-axis has two maxima at $z=\pm 3.7$~\AA, which is in good
agreement with the corresponding value of $z=\pm 3.5$~\AA\ inferred
from spectroscopic measurements.\cite{beck79} The positions of these
two density maxima remain unchanged as $N$ increases to 14.  As
evident in Fig.~\ref{fig:meshden}, the local density near the benzene
impurity is highly structured, which is reflective of the strong
anisotropy in the helium-benzene interaction potential.  We note that
neglecting the rotational terms in the benzene kinetic energy,
Eq.~(\ref{eq:impke}), leads to a considerably more strongly peaked
helium density distribution in the BF frame.\cite{patel02}

\subsection{\label{subsec:excite}Excited states}

\begin{figure}
\resizebox{\columnwidth}{!}{\includegraphics{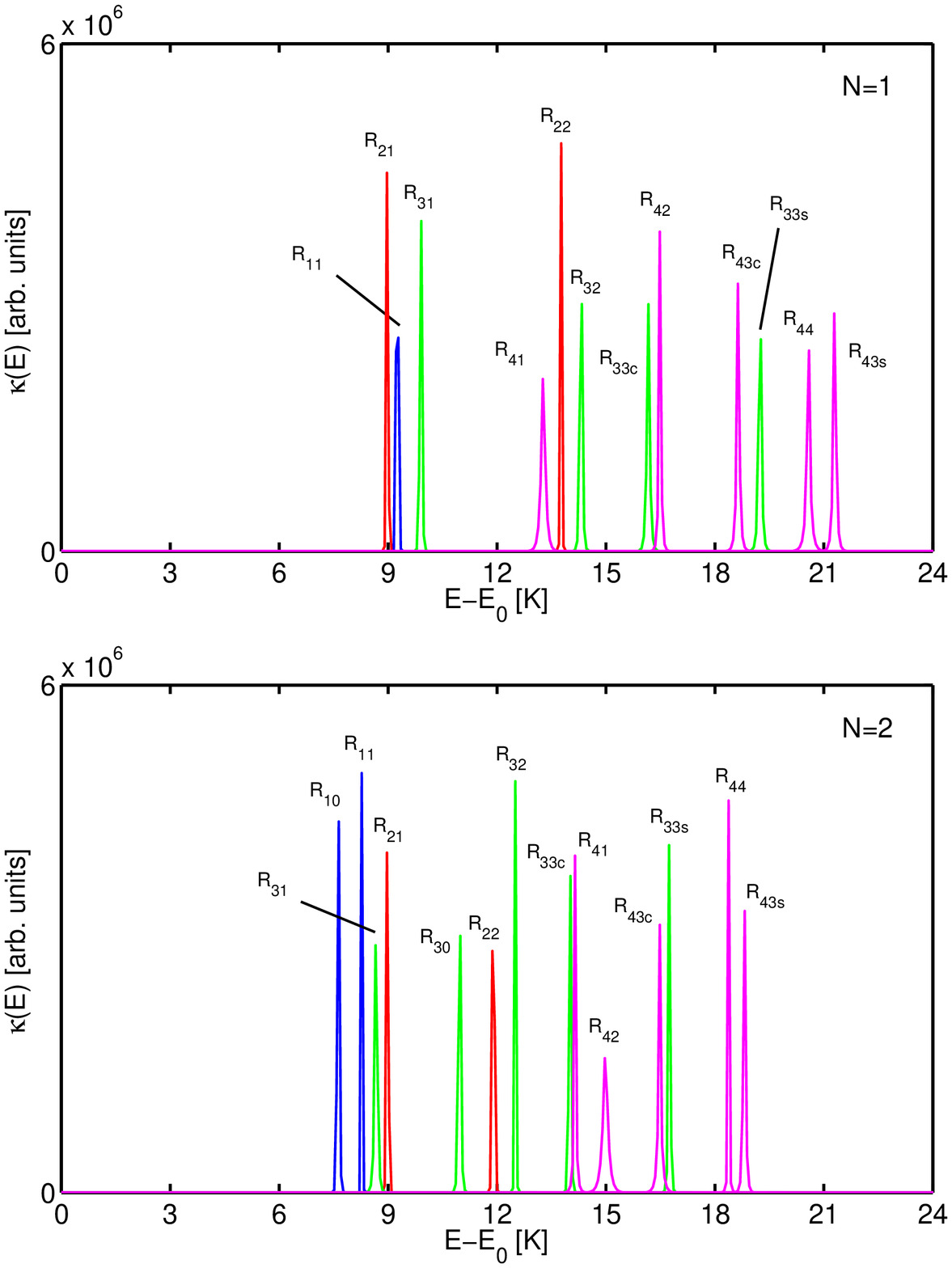}}
\caption{\label{fig:n1n2spec}POITSE spectrum for $N=1$ (upper panel)
and $N=2$ (lower panel).  The peaks are color-coded according to
excitations derived from excitation operators of different $l$
(Table~\ref{tab:proj}).}
\end{figure}

We have calculated helium vibrational excited-state energies for the
$^4$He$_N$-benzene system, using the POITSE methodology described in
Sec.~\ref{subsec:poitse}.  For the $N=1$ dimer, each individual
excitation operator $\hat{A}_{lm}^{\dagger}$ gives a well-defined
excitation, by which we mean a spectral function $\kappa(E)$
consisting of a single, sharp peak.  This indicates that the trial
excited state given by $\hat{A}_{lm}^{\dagger}\Psi_T$ is a good
approximation to a true eigenstate of the system, having
non-negligible overlap with only a single $\phi_n$.  The various
spectral functions are superimposed and shown in the upper panel of
Fig.~\ref{fig:n1n2spec}.
For $N=1$, these helium vibrational modes begin at $\sim 9$~K above
the ground state.  Below this onset, we would expect to see
excitations involving rotational motion of the much heavier benzene
impurity, which lie in the spectral range studied experimentally by
Beck {\em et al.}\cite{beck79} The excitations shown in the upper
panel of Fig.~\ref{fig:n1n2spec} can be grouped into pairs split by
$\sim 0.6-2.3$~K, corresponding to states generated by the application
of pairs of projectors which are symmetric and antisymmetric with
respect to reflection about the benzene molecular plane
(Table~\ref{tab:proj}).  Since the $N=1$ projectors used here all give
well-defined excitations, we conclude that these states represent
symmetric and antisymmetric tunneling pairs.  We are not able to
extract the lowest tunneling excitation given by the application of
the $R_{10}=z$ operator.  This is likely due to its energy being too
close to the ground state $E_0$, and thus its DMC imaginary-time
evolution is too slow relative to the DMC propagation times used for
the excited-state calculations here.  These tunneling energetics are
comparable to those obtained from basis set calculations of the
2,3-dimethylnaphthalene$\cdot$He complex by Bach {\em et
al.}\cite{bach97} These authors found tunneling excitations associated
with the motion of the single complexed helium atom from one side of
the planar aromatic moiety to the other side, with splittings ranging
from $<0.0014$~K for strongly localized states, and up to $4.6$~K for
highly delocalized states.

The Bose-symmetrized versions of the excitation operators used for
$N=1$ give a similar set of well-defined excitations for $N=2$, which
are shown in the lower panel of Fig.~\ref{fig:n1n2spec}.  Thus, we
conclude that these represent single-particle-like excitations, which
is reasonable since the two helium atoms reside in the two equivalent
global minima along the benzene $C_6$-axis, and are well-separated by
the benzene plane.  In general, the $N=2$ excitation energies $E-E_0$
tend to be slightly lower than those for $N=1$, with the onset of
helium vibrational excitations beginning at $\sim 7.5$~K for $N=2$, as
compared to $\sim 9$~K for $N=1$.  Unlike the situation for $N=1$, the
addition of a second particle now allows for a well-defined excitation
of $A_{2u}$ symmetry at 7.63(4)~K, obtained from the $R_{10}=z$
operator.  We note that neglecting the rotational terms in the benzene
kinetic energy, Eq.~(\ref{eq:impke}), alters the energy spectrum
significantly.\cite{huang01}

\begin{figure}
\resizebox{\columnwidth}{!}{\includegraphics{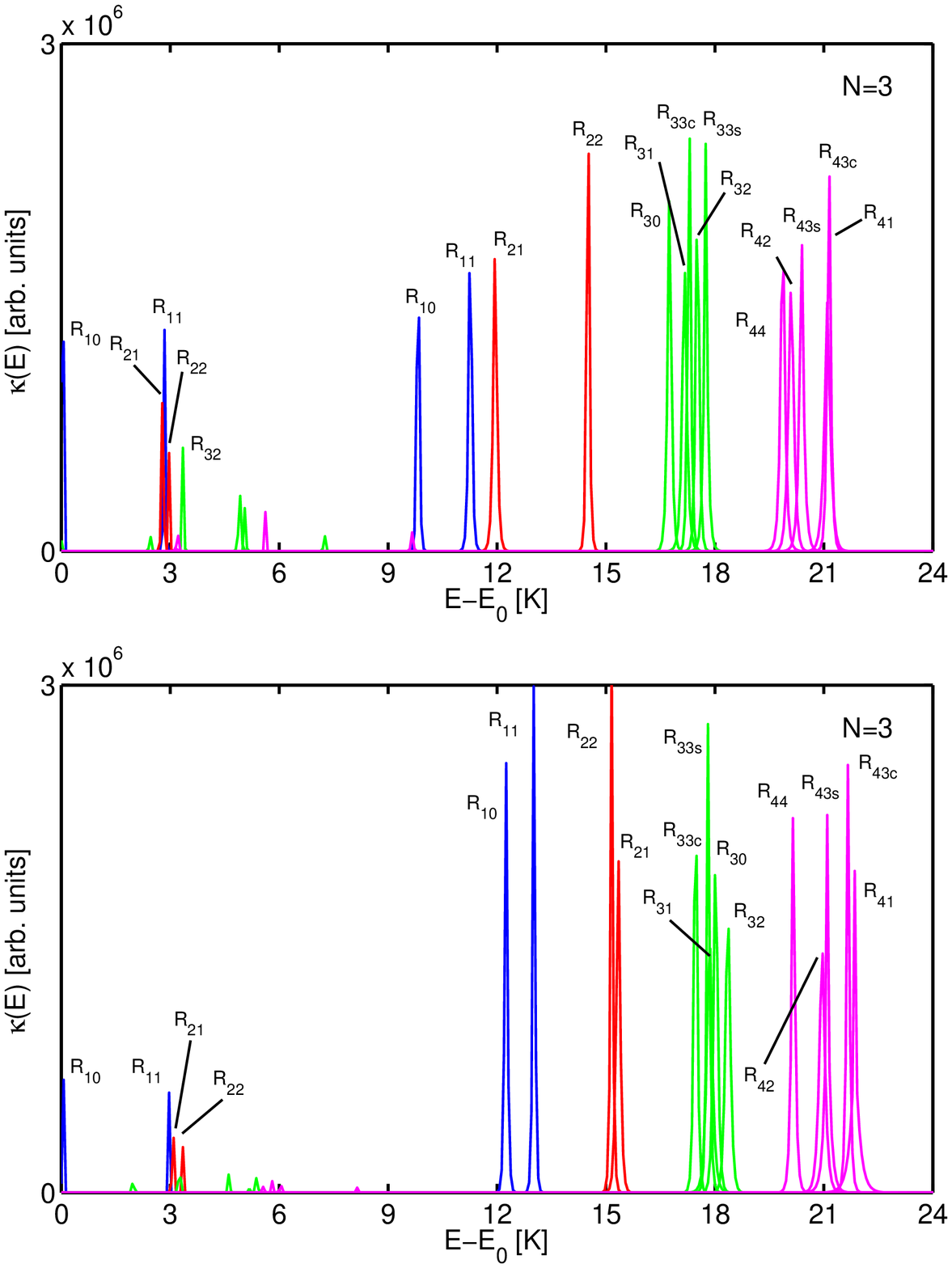}}
\caption{\label{fig:n3spec}POITSE spectrum for $N=3$, deriving from
$\hat{A}^{\dagger}$-type projectors (upper panel) and
$\hat{B}^{\dagger}$-type projectors (lower panel).  The peaks are
color-coded according to excitations derived from excitation operators
of different $l$ (Table~\ref{tab:proj}).}
\end{figure}

With the two global minima occupied at $N=2$, an additional third
helium atom must be delocalized over the twelve secondary minima, due
to the effect of helium-helium repulsions.  Thus, the character of the
excited states is expected to change dramatically as the number of
helium atoms increase from $N=2$ to $N=3$.  This is evident in the
POITSE calculations, where the projectors that give well-defined
excitations for $N=1,2$ (Fig.~\ref{fig:n1n2spec}) now give multiple
peaks for $N=3$, shown in the upper panel of Fig.~\ref{fig:n3spec}.
This indicates that the starting excited state {\em ansatz} generated
by $\hat{A}_{lm}^{\dagger}\Psi_T$ now has appreciable overlap with
more than one eigenstate.  Since the POITSE methodology does not
provide information on excited-state wave functions, interpretation of
the $N=3$ excitations is now less straightforward.  Additional insight
can be gained by making modifications to the projectors.  The most
noticeable new feature of the $N=3$ spectrum here is the appearance of
lower-energy states at $<9$~K.  Since the two atoms situated near the
global potential minima experience a very different local environment
from the third atom that is delocalized over the twelve secondary
minima, the question then arises as to whether one can ascribe any of
the excitations to motion localized near the global potential minima
alone.  We explore this here by defining a set of weighted ``local''
excitation operators
\begin{equation}
\hat{B}_{lm}^{\dagger}(\mathcal{R}) = \sum_{j=1}^N 
\frac{R_{lm}(\mathbf{r}_j)}{\prod_{\gamma\neq\gamma'}\xi_{\gamma}(\mathbf{r}_j)},
\end{equation}
where the product in the denominator is taken over all localization
factors that are {\em not} centered near the global minima.
Application of these $\hat{B}^{\dagger}$-type projectors on $\Psi_T$
gives the following local excited-state {\em ansatz}:
\begin{align}
\Psi & = \hat{B}^{\dagger}_{lm} \Psi_T \\
     & = \sum_{j=1}^N \xi(\mathbf{r}_1) \xi(\mathbf{r}_2) \ldots 
\frac{R_{lm}(\mathbf{r}_j)\xi(\mathbf{r}_j)}{\prod_{\gamma\neq\gamma'}\xi_{\gamma}(\mathbf{r}_j)} \ldots \xi(\mathbf{r}_N) \prod_{i<j}^N \chi(r_{ij}). \label{eq:trial_b}
\end{align}
By substituting Eq.~(\ref{eq:xitrial}) for $\xi(\mathbf{r}_j)$ in
Eq.~(\ref{eq:trial_b}) above, it can be seen that the action of
$\hat{B}^{\dagger}_{lm}$ is to place an atom in an excited
single-particle state that is spatially localized near the set of
sites $\{\mathbf{s}_{\gamma'}\}$ only, {\em i.e.}\ near the two global
potential minima.  These operators are local in the sense that they
act primarily on helium density near these two minima, while still
maintaining spatial and Bose permutation symmetry.  In contrast, by
comparison with Eq.~(\ref{eq:trial_a}), we see that the
$\hat{A}^{\dagger}$ operators are global operators, acting on all
sites.

The spectrum for $N=3$ generated by the set of
$\hat{B}^{\dagger}$-type projectors is shown in the lower panel of
Fig.~\ref{fig:n3spec}.  Compared to the spectrum derived from
$\hat{A}^{\dagger}$-type projectors (upper panel of
Fig.~\ref{fig:n3spec}), the spectral weight of lower-energy states are
now reduced relative to the higher-energy excitations.  A few
low-energy features at $<9$~K persist, in particular the $R_{10}$ and
$R_{11}$ excitations, albeit with reduced weights.  Thus we conclude
that the higher-energy excitations above $\sim 9$~K are associated
with states that are spatially localized near the global minimum of
the interaction potential, while the excitations below this range are
associated with collective helium states of a more delocalized nature.
For each set of states generated by a projector of given $l$,
corresponding states of different $m$ are clustered more closely
together than in the $N=1,2$ spectra, with each cluster of different
$l$ spaced $\sim 3$~K apart.

\begin{figure}
\resizebox{\columnwidth}{!}{\includegraphics{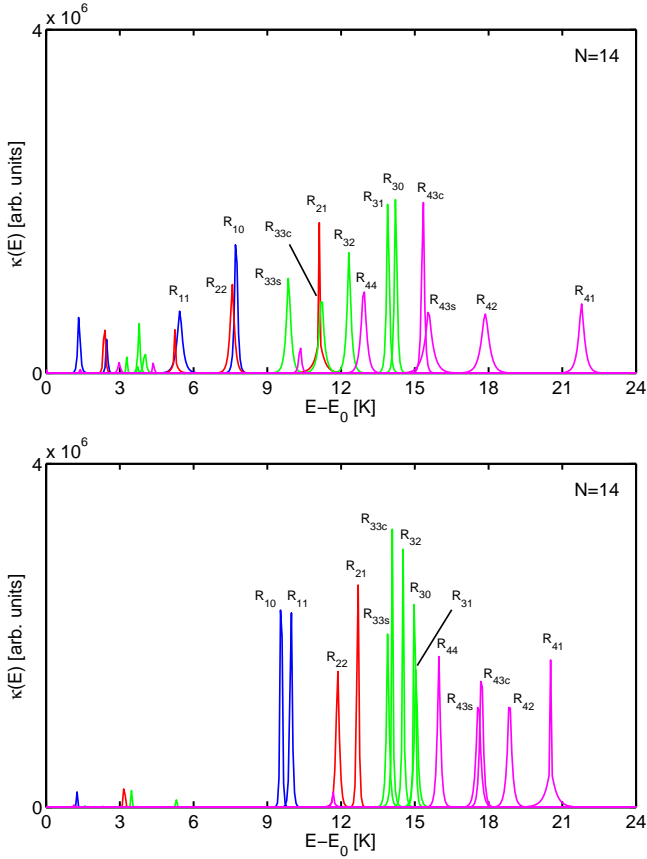}}
\caption{\label{fig:n14spec}POITSE spectrum for $N=14$, deriving from
$\hat{A}^{\dagger}$-type projectors (upper panel) and
$\hat{B}^{\dagger}$-type projectors (lower panel).  The peaks are
color-coded according to excitations derived from excitation operators
of different $l$ (Table~\ref{tab:proj}).}
\end{figure}

As $N$ increases from 3 to 14, the ground-state DMC calculations
reveal that the two global minima and twelve secondary minima are
completely occupied (see Sec.~\ref{subsec:groundstate} and
Fig.~\ref{fig:meshden}).  An unambiguous determination of excited
state energies at this size becomes more difficult due to the increase
in Monte Carlo noise.  However, certain qualitative features remain
apparent.  The upper panel of Fig.~\ref{fig:n14spec}
shows the POITSE excitation spectrum for $N=14$, derived from the
$\hat{A}^{\dagger}$-type global projectors.  These projectors generate
trial excited states which overlap with multiple eigenstates, and the
resulting spectrum for a given projector has multiple peaks at both
low and high energies.  This problem, along with the accompanying
increase in statistical noise in going to $N=14$, gives a spectrum
which is not fully converged in the sense that the peak widths are
broader and the peak positions shift (by up to $\sim 3$~K) with
additional sampling.  Nevertheless, we can again achieve qualitative
insight by comparing with spectra derived from the localized
$\hat{B}^{\dagger}$-type projectors.  The lower panel of
Fig.~\ref{fig:n14spec} shows the spectrum derived from the
$\hat{B}^{\dagger}$-type projectors.  There, the spectral weights of
low-energy excitations ($<9$~K) are again significantly reduced just
as the $N=3$ case, and each projector $\hat{B}_{lm}$ gives a
$\kappa(E)$ consisting of a single dominant peak.  While the
higher-energy $l=4$ excitations are probably still not fully
converged, the $l=1-3$ peaks do not shift much (less than $0.5$~K)
with additional sampling, and appear to be converged.  Thus the
qualitative trends can be clearly established.  Similar to the $N=3$
case, the localized projectors give rise to a set of higher-energy
excitations, except red-shifted by $\sim 3$~K in comparison to the
$N=3$ energies.  For each $l$, states of different $m$ are now
clustered together as was also the case for $N=3$
(Fig.~\ref{fig:n3spec}).  These states represent the localized
vibrational motion of a single helium atom near the global potential
minima, and appear to be associated with the localized helium density
found in the path integral calculations of Ref.~\onlinecite{kwon01}.
The lower-energy excitations, on the other hand, are associated with
collective vibrations delocalized around the periphery of the molecule
surface.

\section{\label{sec:conclusions}Summary Discussion and Implications for larger molecules}

We have computed ground-state energies and structures for small
$^4$He$_N$-benzene clusters, where $N=1,2,3,14$.  For these sizes, the
effect of the strong and highly anisotropic helium-benzene interaction
potential gives rise to a very structured helium density distribution
in the BF frame.  In particular, a single helium atom is
well-localized at each of the two equivalent global potential minima,
above and below the benzene surface.  We find a set of collective
helium excitations with energies of up to $\sim 23$~K above the ground
state.  Among these excitations, the higher-energy states ($>9$~K) can
be characterized as a localized excitation deriving from the helium
density near the global potential minima, {\em i.e.}\ adsorbed on the
molecular nanosurface.  The existence of these localized modes is
consistent with the localization of a single ``dead'' helium atom seen
in Ref.~\onlinecite{kwon01}.  Helium excitations of lower energies
were also obtained, which correspond to collective vibrations of
helium atoms delocalized over equivalent sites of lower binding
energy, situated near the edges of the benzene surface.  Both
localized and delocalized excitations can also be further classified
by their symmetry with respect to the $D_{6h}(M)$ group.

The energetic range of these helium excitations is similar to that
observed experimentally as vibronic structure in the mass-selective
excitation spectra of planar aromatic molecules in small helium
clusters ($N\leq 16$).\cite{even00,even01} It is also similar to the
energetic range of the peaks observed in the phonon wings of vibronic
spectra of planar aromatic molecules in large helium
droplets.\cite{hartmann98,hartmann01,lindinger01,hartmann02} Thus,
these calculations provide support for the picture of helium atoms
adsorbed and vibrating on the molecule surface, with the specific
details of the helium motions being determined by the geometry of the
surface.  In particular, the close relationship between the anisotropy
of the molecule-helium interaction and the nature and energetic range
of the excitations seen here for benzene implies that these
molecule-induced vibrational modes will be very sensitive to the
identity of the molecule, possibly even to the extent of providing a
spectral ``fingerprint'' of complex polyaromatic species.  Since these
calculations have less than one solvation shell of helium surrounding
the molecule, they are directly relevant to the recent experimental
observations for small cluster sizes.\cite{even00,even01} Moreover,
the analysis of the excitations has been made in terms of
single-particle type excitation operators acting on the many-body
ground state, and is thus applicable to any number of solvating helium
atoms.  The fact that the localized excitation associated with the
most strongly bound helium density at the global minimum persists in
the largest cluster studied here ($N=14$), together with the previous
identification of localization at that site in larger helium
clusters,\cite{kwon01} indicates that these localized excitations will
also be present in much larger helium droplets.  For the
$^4$He$_N$-benzene system here we have examined the limit of a single
helium adsorbed atom on a single site given by the benzene
nanosurface.  But given the energetics that we find here, in a more
general sense this class of surface-adsorbed vibrations are very
likely to be responsible for the structure seen in the phonon wing
sideband in experiments using helium droplets.\cite{hartmann02}

A detailed analysis of the experimental phonon wing data clearly
requires molecule-specific calculations with accurate potential
surfaces.  For example, the region of spatial confinement for an
adsorbed helium atom near the global potential minimum is smaller in
benzene than in the larger polyaromatic molecules studied
experimentally with small numbers of helium atoms in
Ref.~\onlinecite{even01}.  In a general context, the present results
are very encouraging in that they show that the diffusion Monte-Carlo
based methodologies can be used to systematically study these
excitations as a function of the number of helium atoms, provided that
realistic molecule-helium interaction potentials are available.  We
emphasize here that a {\em quantitative} analysis, both for the small
cluster vibronic excitations and for the phonon wing structure in
electronic absorption spectra, will require detailed knowledge of the
molecule-helium interaction potential in both ground {\em and}
electronically excited states.

Less obvious than the relation to phonon wing structure is whether the
excitations of the type studied here are responsible for the
splittings on the order of $\sim 1$~K in the zero-phonon line, which
were observed for some large organic impurity molecules in large
helium droplets.  This zero-phonon line splitting has been
experimentally studied in detail for
tetracene,\cite{hartmann01,lindinger01} and for indole
derivatives.\cite{lindinger01} In both experimental studies rotational
fine structure has been ruled out as being responsible and several
different interpretations have been advanced.  For tetracene, a
molecule possessing a plane of mirror symmetry, it has been suggested
that either two inequivalent helium binding sites exist, possibly due
to an inhomogeneous solvation environment, or that some kind of
two-particle tunneling is manifested.\cite{hartmann01} For indole
derivatives, molecules possessing aromatic rings but no mirror
symmetry and not usually completely planar, the theoretical evidence
of localized helium atoms at aromatic rings in larger, superfluid,
helium clusters\cite{kwon01} has been used to propose models of helium
atoms similarly localized on either side of the aromatic portion of
the molecule.\cite{lindinger01}

The $^4$He$_N$-benzene calculations reported here show low-lying
excited states ($\leq 9$~K) for the larger clusters, $N=3,14$.
Furthermore, the spectral weights of these low-lying states do
decrease significantly relative to the higher-energy states when the
local $\hat{B}^{\dagger}$-type operators are applied.  This shows that
the low-energy spectral features we observe the our calculations are
due to vibrational motion of helium located near the periphery of the
benzene surface, since the $\hat{B}^{\dagger}$-type operators
specifically de-emphasize states that are not strongly localized near
the global potential minima.  In contrast to the higher-lying
localized states, it is more difficult to say whether these low-energy
states would remain low-energy with increasing $N$, since addition of
more helium atoms would be expected to significantly change the local
details of the helium wave function in these edge regions around the
molecule.  Thus, it is also more difficult to conclusively claim that
these low-energy delocalized modes are responsible for the
experimentally observed splittings in the zero-phonon lines.

However, with aromatic molecules of lower overall symmetry like the
indole derivatives, it appears reasonable that the effective potential
on opposite sides of the aromatic ring may differ, due to the effect
of non-symmetric, three-dimensional side chains, resulting in slightly
different localized helium densities on either side of the aromatic
ring, as proposed in Ref.~\onlinecite{lindinger01}.  This suggests
that an alternative approach to interpret the splitting of zero-phonon
lines in experiments would be in the context of impurity spectra in
solids, where the impurity molecule can be trapped in structurally
inequivalent trapping sites, giving rise to small differences in the
spectral shifts of the electronic origin.\cite{rebane70} Quantitative
analysis of this kind would also require determination of accurate
helium solvation densities around the molecule.

In extrapolating conclusions made from small cluster studies, whether
theoretical or experimental, to explain experimental observations made
in large droplets, the key point to consider is whether the
localization induced by the helium-aromatic molecule interaction is
sufficiently strong such that the localized helium excitations
characterized here persist with increasing $N$.  As the number of
helium atoms increases, collective compressional and surface modes of
the droplet are expected to develop, and the coupling to these modes
are manifest in the phonon-wing sideband in electronic spectra
measured in large droplets.\cite{hartmann96} The path integral results
for benzene in $^4$He$_{39}$ provide evidence that the single
localized helium atom located on either side of the molecule near the
global potential minimum does indeed retain its localized identity in
large clusters.\cite{kwon01} This provides a strong argument for the
persistence of the single-particle-like localized states whose
excitations have been characterized here for smaller clusters ($N\leq
14$).  An important task for future work is to understand how the
discrete spectra for larger polyaromatics vary with the number of
helium atoms when this is still less than a solvation
shell,\cite{even01} as well as what happens to these excitations in
much larger helium clusters.  A theoretical prerequisite for study of
all these excitations is now the development of accurate trial
functions for these very anisotropic aromatic systems at larger
numbers of helium atoms ($N>14$).  This will then allow analysis of
the transition from localized to collective helium vibrations, as well
as identification of any persistent localized modes that may exist at
an aromatic molecular nanosubstrate.

\section{Acknowledgments}

We acknowledge financial support from the National Science Foundation
under grants CHE-9616615 and CHE-0107541.  Computational support was
provided in part by the National Partnership for Advanced
Computational Infrastructure (NPACI) at the San Diego Supercomputer
Center.  KBW thanks the Miller Institute for Basic Research in Science
for a Miller Research Professorship for 2002--2003.


\end{document}